# Superconducting Transition and Vortex Pinning in Nb Films Patterned with Nano-scale Hole-arrays

U. Welp, Z. L. Xiao, J. S. Jiang, V. K. Vlasko-Vlasov, S. D. Bader, G. W. Crabtree

Materials Science Division, Argonne National Laboratory, Argonne, IL 60439

J. Liang, H. Chik, J. M. Xu

Division of Engineering and Department of Physics, Brown University, Providence, RI 02912

Nb films containing extended arrays of holes with 45-nm diameter and 100-nm spacing have been fabricated using anodized aluminum oxide (AAO) as substrate. Pronounced matching effects in the magnetization and Little-Parks oscillations of the superconducting critical temperature have been observed in fields up to 9 kOe. Flux pinning in the patterned samples is enhanced by two orders of magnitude as compared to unpatterned reference samples in applied fields exceeding 5 kOe. Matching effects are a dominant contribution to vortex pinning at temperatures as low as 4.2 K due to the extremely small spacing of the holes.

The behavior of superconducting vortices in the presence of periodic arrays of pinning sites has attracted much recent attention because of technological as well as scientific interest. Tailored pinning arrays that optimize pinning forces at the theoretical pair-breaking limit in specified field ranges can be designed. In addition, the interplay between the periodic attractive pinning forces and the elastic repulsive vortex-vortex interactions generate a variety of novel static and dynamic vortex phases. In these systems pronounced commensurability effects [1, 2] including large enhancements of the critical current, dips in the magneto-resistance and in the flux creep rate have been observed at fields corresponding to integer multiples or to simple fractions of the matching field, $H_1$. At $H_1$ the number of vortices equals the number of pinning sites. Numerical simulations [3] predict the stable vortex configurations and possible mechanisms of depinning. These configurations have largely been confirmed in direct vortex imaging experiments [4, 5]. Generally, lithographic techniques have been employed to create periodic pinning arrays of various geometries in the form of arrays of micro-holes [1], magnetic dots [2] and irradiation damage [4]. Interestingly, common to all the previous reports on such samples is that commensurability effects are observed only in a very narrow temperature range below the superconducting transition temperature, $T_c$, where pinning forces and critical currents are small even though enhanced pinning is present in the patterned samples at low temperatures when compared to unpatterned reference samples.

Here we show that anodized aluminum oxide (AAO) membranes constitute ideal substrates for growing laterally patterned superconducting films. Large area (several cm$^2$) Nb films containing triangular lattices of 45-nm diameter holes with 100-nm period have been prepared. These samples, having the smallest feature size in extended patterns yet reported, show pronounced commensurability effects even at liquid He-temperature in fields up to 9 kOe. Enhancements of the critical current by more than two orders of magnitude in fields of several kOe have been achieved. In contrast, matching effects in previous studies on Nb-, Pb- or high-$T_c$-samples have been generally limited to fields below 100 Oe [1, 2]. The unique feature of these samples is that the zero-temperature coherence length, (0), and penetration depth, (0), are of the same order as the pattern period ensuring that even at low temperatures the potential energy of the vortices is dominantly periodic. The superconducting – normal phase boundary is characterized by pronounced Little-Parks oscillations in the field dependence of $T_c$, a feature not seen in Nb-, Pb- or high-$T_c$-samples with larger hole spacing prepared lithographically.

The AAO membranes were generated by anodizing high-purity Al foils in an acid solution using a two step process [6]. The Nb films of 100-nm thickness were directly deposited onto the AAO-membranes using magnetron sputtering and were capped with a 10-nm Ag layer. Simultaneously with the Nb/AAO samples continuous reference samples were grown on sapphire substrates.

Figure 1 shows a SEM image of a cleaved edge of the Nb/AAO sample seen under 45 deg. The walls between the pores in the AAO are visible near the bottom of the image revealing a lattice constant of the triangular pore array of a = 101 nm. The Nb film grows on the top of the walls between the holes in a columnar fashion as evidenced by the block-like structures seen at the corner and by boundaries visible in the film surface. The average hole diameter in the resulting Nb-film is d = 45 nm. The holes are perfectly ordered in domains of several μm size.

The resistive transitions in fields between 0 and 7.5 kOe measured in field-cooling are shown in Fig. 2. $T_c$ of the Nb/AAO sample is 6.85 K whereas for the reference sample it is 7.5 K. This reduction in $T_c$ can be caused in part by the proximity with the Ag cap layer and by contamination and/or interdiffusion at the AAO – Nb interface. A remarkable feature in Fig. 2 is the bunching of the transitions at field values labeled $H_1$ to $H_3$. Choosing the resistive mid-points as criterion for $T_c(H)$ yields the superconducting phase diagram as shown in Fig. 3. Since with increasing field the transition shifts in an essentially parallel fashion the features of the phase diagram described below are independent of the resistive criterion. The upper critical field, $H_{c2}$, is characterized by a strongly non-linear temperature dependence near $T_c$ on which an oscillatory variation is superimposed. Around 6 K a cross-over into a linear temperature dependence is observed. For comparison, the conventional linear temperature dependence of $H_{c2}$ measured on the reference film is included in Fig. 3. The non-linear temperature variation of $H_{c2}$ of the perforated film is well described by a parabolic dependence as shown by the dashed line in Fig. 3. Similar results have been observed on superconducting wire networks [7-9] and can be understood in terms of a one dimensional (1 D) nature of superconductivity that occurs at high temperatures when the coherence length becomes larger then the width, w, of the Nb sections between the holes; here w ≈ 55 nm. Then the upper critical field is given by the thin film expression [10] $H_{c2}(T) = \sqrt{12}\ \Phi_0/[2\pi w \xi(T)]$ with $\xi(T) = \xi(0)(1-T/T_c)^{-1/2}$ the Ginzburg-Landau coherence length. A fit (dashed line in Fig. 3) yields $\xi(0) \approx 10$ nm. The same value is obtained from extrapolating the linear $H_{c2}(T)$ dependence observed below 6 K and using the bulk expression for $H_{c2}$, $H_{c2}(T) = \Phi_0/[2\pi \xi(T)^2]$. This value is significantly smaller than the

BCS coherence length of Nb, $\xi_0$ = 38 nm [11], indicating that our films are in the dirty limit with a electron mean free path of $l = 1.38\ \xi(0)^2/\xi_0 \approx$ 5 nm. The Ginzburg-Landau parameter, $\kappa = \lambda/\xi$, can be estimated using the dirty-limit expressions for the penetration depth and the coherence length to $\kappa = 0.72\ \lambda_L/l \approx 7$; $\lambda_L$ = 39 nm is the London penetration depth of Nb [11]. Ratios of $\xi(0)/a \approx 0.1$ and $\lambda(0)/a \approx 0.7$ are achieved which are substantially larger than in previous reports [1,2] on perforated Nb-, Pb- or high-$T_c$-samples. The oscillatory behavior of the phase boundary is seen in detail when plotting the difference, $\Delta T_c$, between the measured $T_c(H)$ and the parabolic fit as shown in the inset of Fig. 3. Cusps in $\Delta T_c$ occurring periodically at multiples of the field $H_1$ = 2.3 kOe are observed. This field value is in very good agreement with the matching field of the triangular unit cell of our pattern, $\Phi_0/[0.5\sqrt{3}\ a^2]$ = 2255 Oe. The observed oscillations are a direct consequence of fluxoid quantization in a multiply connected superconductor and have been studied in superconducting wire networks [7-9], Josephson junction arrays [12] and perforated Al films [13]. Their microscopic origin is identical to those seen in the Little-Parks oscillations [14] of individual superconducting loops: the kinetic energy associated with the supercurrents flowing around the holes in order to maintain fluxoid quantization causes a periodic suppression of $T_c$. $T_c(H)$ for a variety of network topologies has been calculated using linearized Ginzburg-Landau theory [15] which in a phenomenological model [13] can be approximated as a sequence of parabolas: $\Delta T_c = -T_c\ (\xi(0)/a)^2\ [1/4 - (\Phi/\Phi_0 - n - 1/2)^2]$. Here, $\Phi = 1/2\ \sqrt{3}\ a^2\ H$ is the flux per unit cell, and n = 0, 1, 2, … is the number of flux quanta per cell. In contrast to the behavior of networks, though, perforated superconducting samples offer the possibility of interstitial vortices residing in the superconducting segments between the holes. However, since in the field range where the oscillations occur, that is, below 8 kOe, the coherence length at the onset of superconductivity, $\xi(T_{c2})$, is larger than the width of the Nb-segments as evidenced by the 1D variation of $H_{c2}$ vortices do not nucleate in these sections. Then the supercurrents are governed by the net flux per unit cell, and the above relation is applicable. As the temperature decreases, the coherence length becomes shorter than the width of the Nb sections and interstitial vortices may appear. The low-temperature vortex configuration will be determined by the interplay of the energies of interstitial vortices and multi-quanta vortices located in the holes [16] (see below).

A fit of the oscillatory $T_c$-variation to the above expression for the states with n = 0, 1, 2, is shown in the inset of Fig. 3 as dotted lines. The data are well described, and a value of 15 nm for the coherence length is obtained, which is in reasonable agreement with the above determination based on the $H_{c2}$-line. Deviations from the fit occur notably near the bottom of the n = 0 – parabola. These are indicative of additional cusps in the $T_c(H)$ dependence occurring at fractional filling factors [7-9, 13, 15] which are not accounted for in above model. The amplitude of the $T_c$-oscillations of about 50 mK is close to the expected value of $T_c(\xi(0)/a)^2 \approx 10^{-2} T_c$.

Figure 4 shows magnetization loops on a linear scale (top) and the width, $\Delta m$, of the loops on a log-scale (bottom) of a square shaped sample with sides L = 2 mm. Also included are $\Delta m$ - data on a continuous reference sample of the same shape as the perforated film. Pronounced step-wise variations of the magnetization occur at multiples of the matching field, extending the useful range of periodic pinning to above 5 kOe. In contrast to the behavior seen in previous reports [1, 2] matching is the dominant feature in the magnetization curves even at the lowest temperatures. In comparison to the continuous film, flux pinning in the Nb/AAO sample is dramatically enhanced - by more than two orders of magnitude in the field range of several kOe. (Note, that $T_c$ of the unpatterned sample is 0.7 K higher than that of the Nb/AAO sample).

The observation of matching phenomena at fields $H_n = n H_1$ with $n > 1$ can be explained in terms of caging of interstitial vortices between the holes [17] or in terms of multi-quanta vortices residing in the holes [18]. These considerations, formulated in the London approximation, apply at sufficiently low temperatures when the coherence length becomes shorter than the hole diameter and the width of the intervening segments. For a triangular array of cylindrical cavities it has been shown [18] that doubly quantized vortices are stable for $d^3 > 8 \; a^2$. This condition is not fulfilled for our samples, which indicates the presence of single quanta vortices in the entire low - temperature range. Similarly, for an isolated hole it has been shown [19] that the attractive pinning force vanishes if the number of vortices confined to the cavity exceeds the saturation number $n_s = d/4\xi(T)$, which yields $n_s = 1$ for our samples. Thus, at low temperature when the coherence length becomes shorter than the width of the Nb segments single-quanta vortices coexisting with interstitial vortices are the stable configuration. In addition, the magnetization data in Fig. 4 taken mostly at temperatures below the 1D – 3D cross-over of the $H_{c2}(T)$-line represent flux-gradient-driven states in which, in contrast to field-cooled states, multi-quanta vortices may never appear. The interstitial vortices are pinned by a confining cage potential exerted by vortices trapped in the holes. This potential is determined by the ratio $\lambda(T)/a$ and the number of interstitials per unit cell [17]. As $\lambda(T)/a$ decreases, this potential is rapidly smoothed out, and the periodic pinning potential for interstitial vortices disappears. In our samples $\lambda(T)/a \gtrsim 1$ assuring that the periodic caging potential is a dominant contribution to the pinning force over the entire temperature range. In arrays with periods on the µm-scale additional pinning due to the elastic interactions between the vortices trapped in the holes and the interstitial vortices is still expected. However, since the anchored vortices are spaced far apart, a >> (T) at low temperatures, their periodic arrangement does not contribute significantly to the total vortex energy.

As the number of interstitial vortices increases their interaction energy becomes an important contribution to the total energy and matching effects decrease. In our samples matching effects for n > 3 are strongly suppressed even at the lowest temperature (Fig. 4b). This is consistent with the presence of two natural interstitial sites per unit cell as indicated in the schematic in Fig. 4. At higher filling factors, less stable and thus less strongly pinned configurations arise.

The magnetization data yield no indication of pinning effects at fractional matching fields. This may arise since in the flux-gradient driven dynamics the strong single vortex pinning at the holes precludes the formation of ordered superlattices which would minimize the vortex-vortex interaction energy whereas in field-cooled measurements (Figs. 2, 3) these equilibrium states may develop. The strongly non-monotonic low-field dependence of the magnetization observed at 4.5 K is not due to matching effects but is caused by flux avalanches. This behavior is frequently observed in high-pinning superconductors [20] and has been confirmed directly using magneto-optical imaging [21].

In conclusion, we have shown that anodized aluminum oxide membranes are ideal patterned substrates for fabricating perforated superconducting films. Large area (several

cm$^2$) Nb films containing triangular lattices of 45-nm diameter holes with 100-nm period have been prepared. These samples have the smallest feature size in extended patterns yet reported, and show pronounced commensurability effects down to 4.2 K in fields up to 0.9 T. Enhancements of the critical current by more than two orders of magnitude in fields of several kOe have been achieved. The unique feature of these sample is that the zero-temperature coherence length, (0), and penetration depth, (0), are of the same order as the periodic pattern ensuring that even at low temperatures the periodic vortex potential is dominant.

This work was supported by the U.S. Department of Energy, BES, Materials Science under contract W-31-109-ENG-38 at Argonne National Laboratory, and by the National Science Foundation under contract NSF ECS 0070019 and by the Office of Naval Research under contract N00014 001 0260 at Brown University. The SEM work was performed at the Electron Microscopy Center at ANL.

References

[1] A. T. Fiory *et al*., Appl. Phys. Lett. **32**, 73 (1978); M. Baert *et al*., Phys. Rev. Lett. **74**, 3269 (1995); M. Baert *et al*., Europhys. Lett. **29**, 157 (1995); V. V. Moshchalkov *et al*., Phys. Rev. B **54**, 7385 (1996), *ibid.* **57**, 3615 (1998); E. Rosseel *et al*., Phys. Rev. B **53**, R2983 (1996); A. Castellanos *et al*., Appl. Phys. Lett. **71**, 962 (1997); V. V. Metlushko *et al*., Europhys. Lett. **41**, 333 (1998); V. V. Metlushko *et al*., Phys. Rev. B **59**, 603 (1999), *ibid.* **60**, R12585 (1999); L. Van Look *et al*., *ibid.* **60**, R6998 (1999).

[2] J. I. Martin *et al*., Phys. Rev. Lett. **79**, 1929 (1997), *ibid.* **83**, 1022 (1999); D. J. Morgan, J. B. Ketterson, *ibid.* **80**, 3614 (1998); Y. Jaccard *et al*., Phys. Rev. B **58**, 8232 (1998); A. Hoffmann et al., *ibid.* **61**, 6958 (2000); A. Terentiev *et al*., *ibid.* **61**, R9249 (2000); O. M. Stoll *et al*., cond-mat/0108321.

[3] Ch. Reichhardt et al., Phys. Rev. B **57**, 7937 (1998); *ibid* **63**, 054510 (2001);.*ibid* **64**, 052503 (2001); *ibid* **64**, 144509 (2001); V. I. Marconi, D. Dominguez, Phys. Rev. B **63**, 174509 (2001); B. Y. Zhu et al., Phys. Rev. B **64**, 012504 (2001).

[4] K. Harada et al., Science **274**, 1167 (1996).

[5] S. B. Field *et al*., Phys. Rev. Lett. **88**, 067003 (2002); A. N. Grigorenko *et al*., Phys. Rev. B **63**, 052504 (2001); D. M. Silevitch *et al*., J. Appl. Phys. **89**, 7478 (2001).

[6] A. J. Yin *et al*., Appl. Phys. Lett. **79**, 1039 (2001); J. Liang et al., J. Appl. Phys. **91**, 2544 (2002); H. Masuda, H. Fukuda, Science **268**, 1466 (1995).

[7] B. Pannetier et al., Phys. Rev. Lett. **53**, 1845 (1984).

[8]  V. V. Moshchalkov et al., Nature **373**, 319 (1995); V. Bruyndoncx *et al*., Europhysics Lett. **36**, 449 (1996); T. Puig *et al*., Phys. Rev. B **58**, 5744 (1998).

[9]  M. Tinkham et al., Phys. Rev. B **28**, 6578 (1983); C. W. Wilks et al., Phys. Rev. B **43**, 2721 (1991); H. S. J. van der Zant et al., Phys. Rev. B **50**, 340 (1994).

[10] M. Tinkham, "Introduction to Superconductivity" (McGraw-Hill, New York, 1996), Ch. 4.10.

[11] B. W. Maxfield, W. L. McLean, Phys. Rev. **139**, A1515 (1965).

[12] W. J. Elion, H. S. J. Van der Zant, J. E. Mooij, "Macroscopic Quantum Phenomena and Coherence in Superconducting Networks" (World Scientific, Singapore, 1995).

[13] A. Bezryadin, B. Pannetier, J. Low Temp. Phys. **98**, 251 (1995).

[14] R. D. Parks, W. A. Little, Phys. Rev. A **133**, 97 (1964).

[15] J. Simonin et al., Phys. Rev. Lett. **49**, 944 (1982); H. J. Fink et al., Phys. Rev. B **26**, 5237 (1982); R. Ramal et al., Phys. Rev. B **27**, 2820 (1983); S. Alexander, Phys. Rev. B **27**, 151 (1983); F. Nori et al, Phys. Rev. B **36**, 8338 (1987).

[16] A. Bezryadin, B. Pannetier, J. Low Temp. Phys. **102**, 73 (1996).

[17] I. B. Khalfin, B. Ya. Shapiro, Physica C **207**, 359 (1993).

[18] A. I. Buzdin, Phys. Rev. B **47**, 11416 (1993).

[19] G. S. Mkrtchyan, V. V. Shmidt, Soviet Phys. JETP **34**, 195 (1972).

[20] P. S. Schwartz, C. P. Bean, J. Appl. Phys. **39**, 4991 (1968); R. G. Mints, A. L. Rakhmanov, Rev. Mod. Phys. **53**, 551 (1981); E. R. Nowak et al., Phys. Rev. B **55**, 11702 (1997); P. Esquinazi et al., Phys. Rev. B **60**, 12454 (1999).

[21] V. K. Vlasko-Vlasov et al, unpublished.

Figure Captions

Fig. 1    SEM image of a cleaved edge of the sample viewed under an angle of 45 deg.

Fig. 2    Resistive transitions of the sample measured on field-cooling in fields from 0 to 7.5 kOe every 250 Oe. The bunching of the transitions in field ranges corresponding to $H_1$, $H_2$ and $H_3$ is seen.

Fig. 3    Superconducting phase diagram of the Nb/AAO sample (solid symbols) and of a continuous reference sample (open symbols) as determined from the resistive transitions. The dashed line is a fit of the 1D expression of $H_{c2}(T)$ to the high temperature data. The inset shows the oscillatory variation of $T_c(H)$ taken as the difference between the measured value and that obtained from the parabolic fit. The first three matching fields are indicated. The dotted lines are fits based on a superconducting wire network.

Fig. 4    Field dependence of the magnetization (top) and of the magnetization hysteresis (bottom) of the Nb/AAO sample at various temperatures. Hysteresis data on a continous reference sample (open symbols) are included. The inset shows a schematic of the sample and the most favorable locations of interstitial vortices.

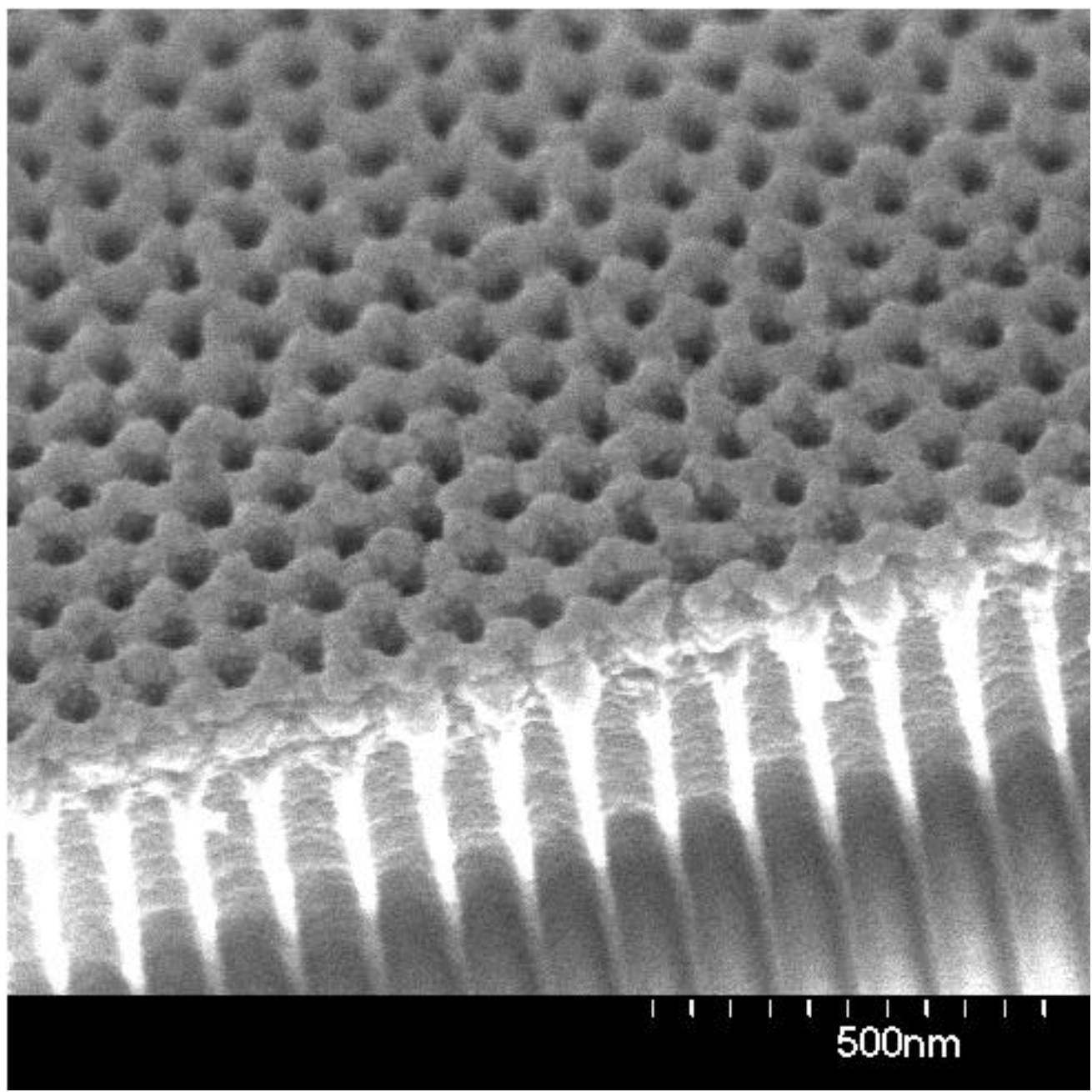

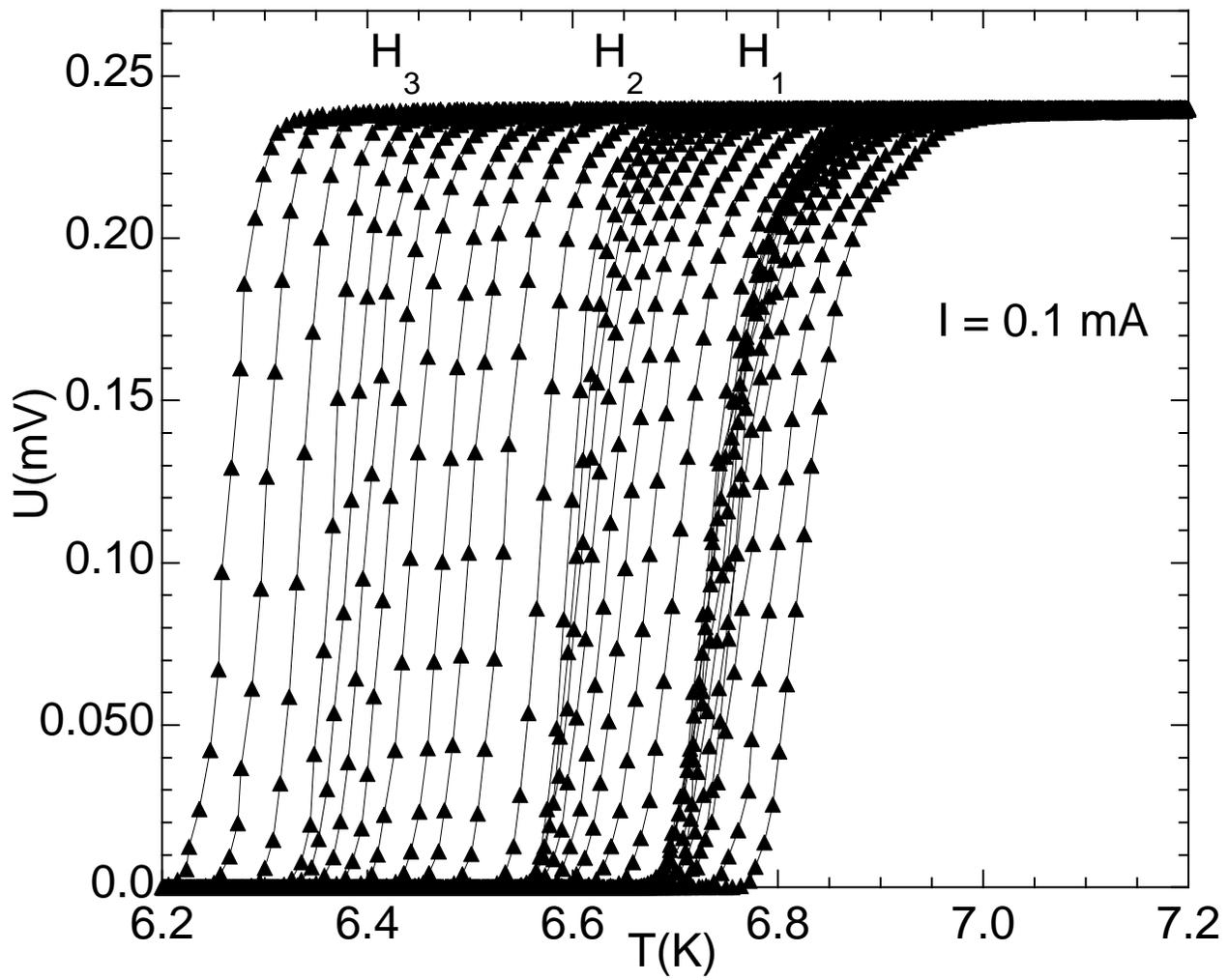

Fig. 2
U. Welp et al.

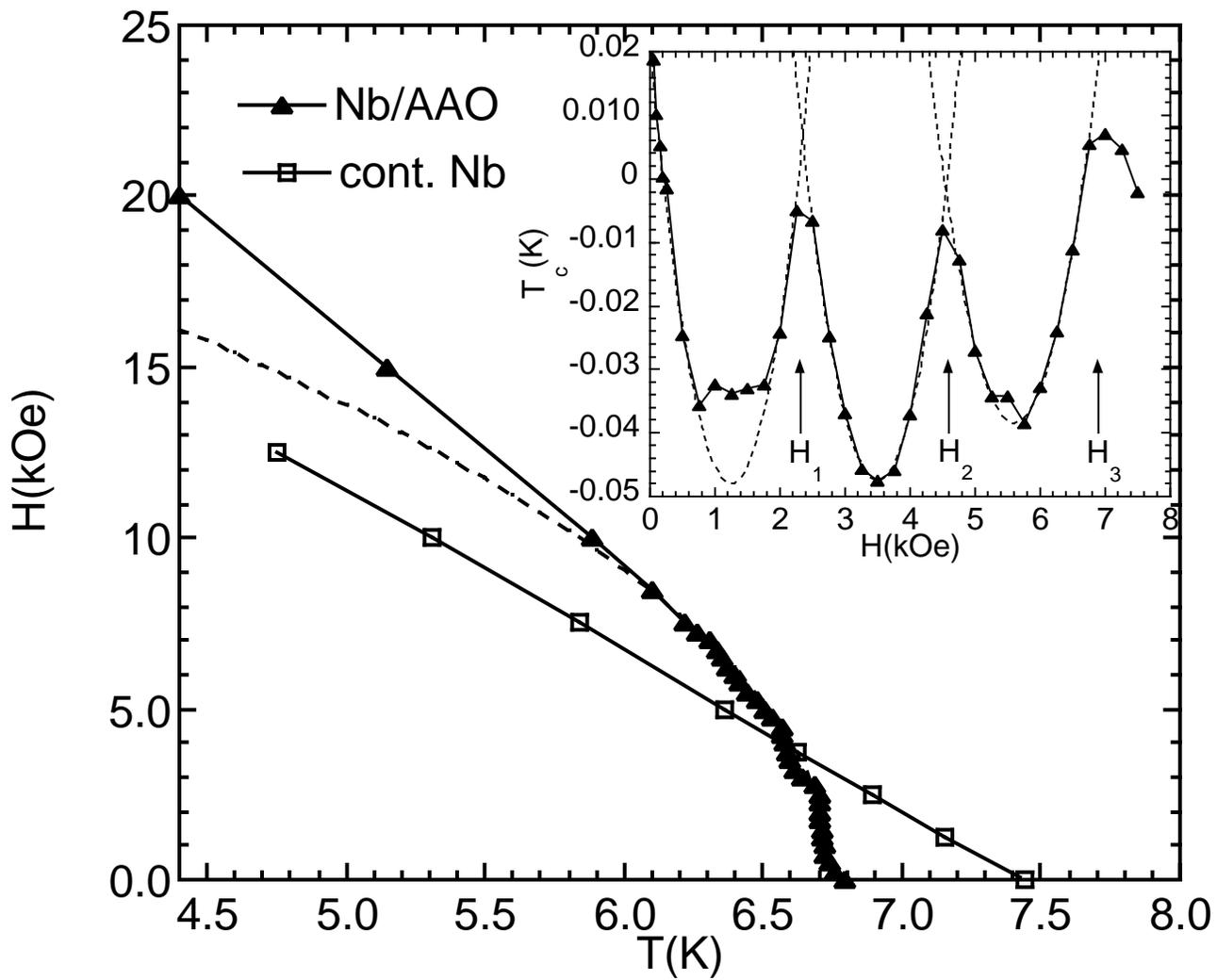

Fig. 3
U. Welp et al.

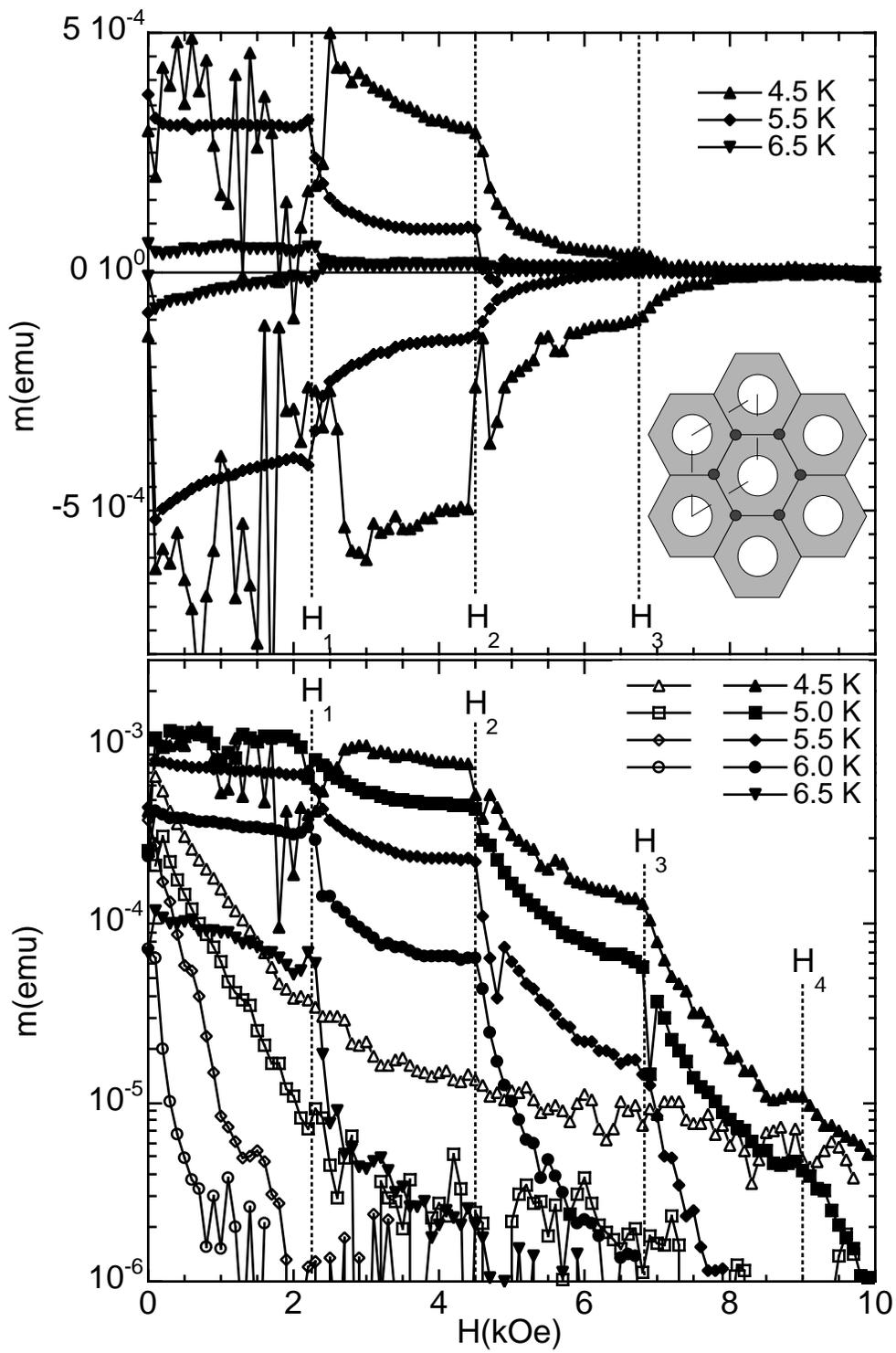

Fig. 4
U. Welp et al.